\documentclass[prb,showpacs,preprintnumbers,amsmath,amssymb,letterpaper,twocolumn]{revtex4}

\usepackage{graphicx}
\usepackage{dcolumn}
\usepackage{bm}
\usepackage{epstopdf}
\usepackage{latexsym}
\usepackage{epsfig}

\newcommand{\be}{\begin{equation}}
\newcommand{\ee}{\end{equation}}
\newcommand{\bea}{\begin{eqnarray}}
\newcommand{\eea}{\end{eqnarray}}

\renewcommand{\Im}{\mathrm{Im}}

\newcommand{\eref}[1]{Eq.~(\ref{#1})}

\newcommand{\ocite}[1]{Ref.~\onlinecite{#1}}

\begin{document}

\title{Fermi-liquid effects in the gapless state of marginally thin superconducting films}

\author{G. Catelani}
\affiliation{Department of Physics and Astronomy, Rutgers University, Piscataway, New Jersey 08854, USA}

\author{X. S. Wu}
\altaffiliation[Present address:]{School of Physics, Georgia Institute of Technology, Atlanta,
Georgia 30332, USA}
\author{P. W. Adams}
\affiliation{Department of Physics and Astronomy, Louisiana State University, Baton Rouge,
Louisiana 70803, USA}

\date{\today}

\begin{abstract}

We present low temperature tunneling density-of-states measurements
in Al films in high parallel magnetic fields. The thickness range of
the films, $t=6-9$ nm, was chosen so that the orbital and Zeeman
contributions to their parallel critical fields were comparable.  In
this quasi-spin paramagnetically limited configuration,  the field
produces a significant suppression of the gap, and at high fields
the gapless state is reached. By comparing measured and calculated
tunneling spectra we are able to extract the value of the
antisymmetric Fermi-liquid parameter $G^0$ and thereby deduce the
quasiparticle density dependence of the effective parameter
$G^0_\mathrm{eff}$ across the gapless state.

\end{abstract}

\maketitle

\section{Introduction}
\label{sec:intro}

One of the most successful constructs in condensed matter physics is
Landau's Fermi-liquid (FL) description of the low temperature
properties of interacting electron systems.
\cite{Landau1956,Landau1957}  The theory provides for the salient thermodynamic and transport
properties of a profoundly complex many-body system in terms of a
low-density gas of quasiparticle excitations.  The averaged effects
of electron-electron and electron-phonon interactions are
incorporated into a number of FL parameters, which are associated
with the renormalization of fundamental system properties such as the
effective mass and the spin susceptibility.\cite{Baym1991}
In particular, the so-called antisymmetric FL parameter $G^0$ affects
the spin response of the system and is related to the
ratio of the spin susceptibility density of states $N(\chi)$ to the
heat-capacity density of states $N(\gamma)$ by
$G^0=N(\gamma)/N(\chi)-1$.\cite{Baym1991} In this paper we report low-temperature
measurements of the parameter $G^0$ in superconducting disordered Al films.

Although $G^0$ is a fundamentally important parameter in the many-body
description of metals, there have been very few measurements of it
reported in the literature.  This is due, in part, to the fact that
it is difficult to measure it directly in bulk systems.
Consequently, little is known about its dependence on
disorder, quasiparticle density, and/or dimensionality.
Measurements of $G^0$ have been extracted from high-field tunneling density of
states (TDOS) and critical-field studies of low atomic mass superconducting
films. For example, estimates of normal-state value of $G^0$
have been reported via tunneling studies of the Zeeman splitting of the
BCS density of states in superconducting Al films, $G^0\sim0.3-0.4$,
\cite{Alexander1985,Tedrow1984} in the intermediate to high-temperature regime where
the superconducting order parameter is partly suppressed by thermal fluctuations.
Low-temperature TDOS measurements in amorphous Ga films, which are
strong-coupling superconductors, give a somewhat larger value of $G^0\sim 0.81$.\cite{Gibson1989}
Alternatively, $G^0$ can be extracted from parallel critical field
measurements.  This method gives $G^0\sim0.23$ in TiN films.\cite{Suzuki2000}
The only direct measurements of $G^0$ have been
obtained in the normal state, via the field dependence of the pairing resonance
\cite{Aleiner1997} in Al, $G^0\sim0.17$,\cite{Butko1999} and Be
films, $G^0\sim0.21$.\cite{Adams2000} In all these systems the
spin-orbit scattering is quite low; thus, spin remains a good quantum
number.

In the experiments described above the films were sufficiently thin so as to suppress the Meissner
currents; therefore the field response was purely due to the electron Zeeman splitting (see e.g.
\ocite{Fulde1973} and Sec.~\ref{sec:phase}).
By contrast, in this work we consider thicker films; this enables us to explore the
gapless state and determine $G^0$ from low temperature TDOS measurements.
In the normal state, the internal
field $H_i$ felt by the electron spins is given by $H_i=H+H_\mathrm{ex}$, where $H$ is the applied field
and the exchange field is $H_\mathrm{ex}=-HG^0/(1+G^0)$.
Deep in the
superconducting phase, {\it i.e}, at low temperature and low field, the
number of quasiparticles is small due to the fact that they have been
consumed by the formation of the superconducting condensate. In
this limit, the exchange effects are greatly suppressed and the effective Landau parameter
[see \eref{Geff}] is small compared to its normal state value,
$G^{0}_\mathrm{eff}\sim0$.
However, as one
approaches the critical field $H_{c||}$, the quasiparticle
density increases and $G^{0}_\mathrm{eff}\rightarrow G^0$.  Below we show that the main features of
the TDOS in the gapless superconducting state are strongly influenced by the rise of these
exchange interactions.
In addition to its fundamental interest, a thorough understanding of exchange effects
in superconducting Al films is important for possible applications in mesoscopic
hybrid structures aimed at the control of current spin polarization via the Zeeman
splitting in the superconducting elements.\cite{Giaz}

This paper is organized as follows: in Sec.~\ref{sec:phase} we give a brief overview of
the phase transition in thin films and introduce the parameters necessary for the description
of its properties. In Sec.~\ref{sec:exp} we give some details of the samples' preparation,
and in Sec.~\ref{sec:res} we discuss our main results.

\section{Phase transition in marginally thin films}
\label{sec:phase}

In bulk superconductors the critical-field
transition is completely dominated by the orbital response of the
conduction electrons.\cite{Tinkham1996} It is possible,
however, to inhibit the orbital response by applying a magnetic
field in the plane of a superconducting film whose thickness $t$ is much
less than the superconducting coherence length $\xi$ and whose electron
diffusivity is low.\cite{Fulde1973,Meservey1970,Wu1994}  Under these conditions
the phase transition to the normal state is
mediated by the spin polarization of the electrons, and at the critical field the
electron Zeeman splitting is of the order of the superconducting gap
energy. At zero temperature, in particular,
a first-order transition from the superconducting state to the
paramagnetic normal state occurs at the Clogston-Chandrasekhar
critical field\cite{Clogston1962,Chandrasekhar1962}
\begin{equation}
H^{CC}_{c||}=\frac{\Delta_o\sqrt{1+G^0}}{\sqrt{2} \mu_B} \, ,
\label{Clogston}
\end{equation}
where $\Delta_o$ is the zero-field, zero-temperature gap energy, and
$\mu_B$ is the Bohr magneton.
This so-called ``spin-paramagnetic'' (S-P) transition is realized in low atomic mass
superconductors such as Al, Be, and TiN, and it is accompanied by
very peculiar dynamics.\cite{Fulde1973,Wu1995a,Butko1999,Suzuki1984,WAC}
Note that for positive $G^0$ the parallel critical field exceeds the non-interacting
Clogston-Chandrasekhar limit.

In the thin-film limit and at low temperatures, the parallel magnetic field has
little effect on the order parameter up to the transition, and the Clogston-Chandrasekhar
critical field [\eref{Clogston}] can be more than an order of magnitude higher than
its orbitally-mediated counterpart. However, if the thin film
condition is marginally relaxed then orbital effects will generally
lower the critical field.  A measure of the relative weight of the orbital
response compared to the spin polarization in determining the parallel critical field is given by
the dimensionless orbital pair-breaking parameter\cite{Fulde1973}
\begin{equation}
c=\frac{D(et)^2\Delta_o}{6\hbar\mu_{B}^{2}} f\left(\ell/t\right) \, ,
\label{Orbital}
\end{equation}
where $D$ is the electron diffusivity, $t$ is the film thickness,
$e$ is the electron charge, and $\ell$ is the mean-free path for impurity scattering.
The function $f(\ell/t)$ describes the crossover between local electrodynamic response
of the film for $\ell \ll t$, where $f(\ell/t)\to 1$, and the non-local one in the opposite limit
$\ell \gg t$, where $f(\ell/t)\simeq 3t/4\ell$.\cite{Maki}
For marginally thin films used in this study, $c\sim 1$.

Beside the film's thickness (through the parameter $c$)
and exchange effects, spin-orbit scattering also affects
the value of the parallel critical field. Therefore, following \ocite{Fulde1973},
we introduce a third dimensionless parameter
$b=\hbar/3\Delta_o \tau_{so}$, where $1/\tau_{so}$ is the spin-orbit scattering
rate. As remarked in Sec.~\ref{sec:intro},
in the materials under consideration spin-orbit scattering is a small effect, $b\ll 1$.
The three parameters $c$, $G^0$, and $b$, can have competing effects on the value of the
parallel critical field. Moreover, they also affect the position of the
tricritical point separating the low-temperature first-order phase transition from the
higher-temperature second-order one. In the limit $G^0,c \to 0$ the tricritical temperature
is $T_{tri} \simeq 0.56 T_c$, where $T_c$ is the critical temperature.\cite{Fulde1973} More
generally, the tricritical temperature and other properties of a superconducting film, such as the TDOS, can
be calculated by solving the Usadel equations for the semiclassical
Green's functions together with the self-consistent
equations for the order parameter $\delta$ and the internal magnetic
field $H_i$; these equations can be found in Ref.~\onlinecite{Alexander1985}
(see also \ocite{Suzuki2000} for an alternative parametrization).
All three dimensionless parameters enter into these mean field equations, while
$\Delta_o$ (or $T_c \simeq 0.57 \Delta_o/k_B$) fixes the overall energy scale.

In this work we present low
temperature measurements of TDOS
in Al films whose thicknesses are about two to three times greater
than the $t\sim3$ nm typically used in spin-paramagnetic studies.\cite{Butko1999a}
Though the thickness of the films remains much
smaller than the superconducting coherence length, $t\ll\xi$, the orbital
response to the applied field is no longer negligible.  Indeed, the
thickness range is chosen to assure that the orbital and Zeeman
contributions to the critical field are comparable (this could also
be accomplished by rotating the films out of parallel orientation,
but a finite perpendicular field component would introduce vorticity\cite{Wu1995b}).
Interestingly, the
critical field transition can become reentrant\cite{Fulde1967}
under these conditions. The reentrance is associated with a high-entropy
gapless state near the critical field, in which the
superconducting state has higher entropy that the normal state.\cite{Wu2006}
The primary purpose of the present work is to determine
$G^0$ through TDOS measurements across the gapless region, where there is a monotonic increase
in quasiparticle density as one approaches the second-order phase
transition to the normal state.

\section{Sample preparation}
\label{sec:exp}

Aluminum films were fabricated by e-beam deposition of 99.999\% Al
onto fire polished glass microscope slides held at 84~K. The
depositions were made in a typical vacuum $P<3\times10^{-7}$~Torr at a
rate of $\sim0.2$~nm/s. Films with thicknesses
ranging from 6 to 9~nm had normal-state sheet resistances that
ranged from $R=10$ to $20\,\Omega$ at 100~mK. After deposition, the
films were exposed to the atmosphere for 0.5-4~h in order to
allow a thin native oxide layer to form. On top of the oxide, serving
as the tunneling barrier, a 14-nm-thick Al
counterelectrode was deposited. Due to its relatively large thickness,
the counterelectrode had a parallel critical field of $\sim1.1$~T,
which is to be compared with $H_{c\parallel}\sim3$~T for the films.
The junction resistance ranged from $10$ to $20$~k$\Omega$, depending on exposure
time and other factors, for a junction area of about 1$\times$1~mm$^2$.
Only junctions with resistances much greater
than that of the films were used.
Magnetic fields of up to 9 T were applied using a
superconducting solenoid. An \textit{in situ} mechanical rotator was
employed to align the film surface parallel to the field with a
precision of $\sim0.1^{\circ}$. Measurements of resistance and tunneling
were carried out in an Oxford dilution refrigerator using a standard ac
four-probe technique.

\section{Results and discussion}
\label{sec:res}

\begin{figure}
\includegraphics[width=.5\textwidth]{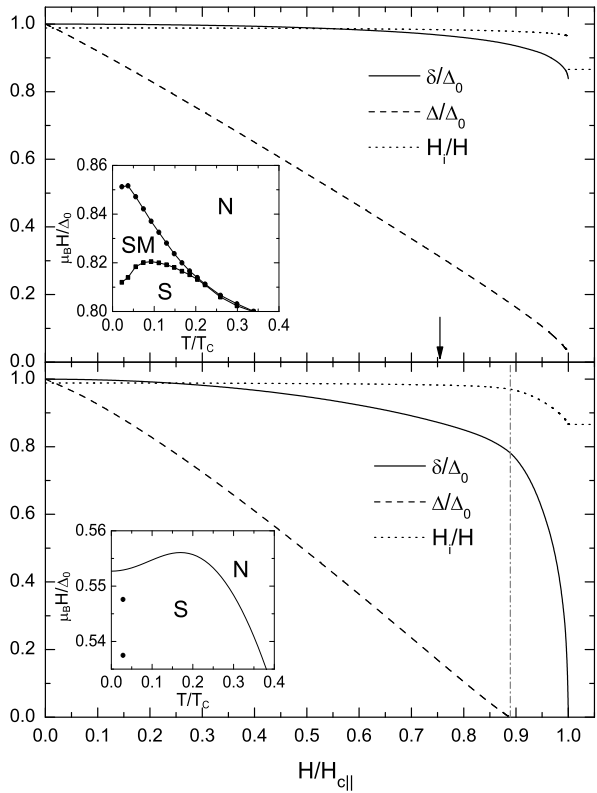}
\caption{\label{OrderParameter}Upper panel: Calculated field
dependencies of the superconducting order parameter $\delta$, gap
$\Delta$, and ratio between internal and applied fields $H_i/H$ for a 2.5-nm-thick Al film
($T_c=2.7$~K) in parallel field at 77~mK. The arrow indicates the theoretical
supercooling field. Note that the gap is finite up to the superheating field.
The upper inset is the
corresponding phase diagram where S is the superconducting phase, N
the normal state, and SM is the state memory region.
Lower panel:
Field dependencies of the same quantities for a 7-nm-thick Al film
($T_c=2.1$~K) at 60~mK.  The vertical dot-dashed line separates the
gapped and gapless regions. Note in the latter the fast drop of the order parameter
with increasing field.
The lower inset is the corresponding
phase diagram, with a maximum in the second-order
critical-field curve. The two points in the phase diagram
represent the temperature and fields at which tunneling spectra were
taken in the gapless region.}
\end{figure}

Shown in the upper panel in Fig.~\ref{OrderParameter} are the
calculated low-temperature field dependencies of the order parameter
$\delta$, gap energy $\Delta$, and internal field $H_i$ for a
2.5-nm-thick Al film having $T_c=2.7\,\mathrm{K}$, $\Delta_o/e=0.41\,\mathrm{mV}$,  $\xi\sim15\,\mathrm{nm}$,
and sheet resistance $R\sim1\,\mathrm{k}\Omega$.   At this thickness and
resistivity orbital effects are negligible, since we estimate $c\simeq 0.02 \ll 1$. Accordingly,
$\delta$ and $H_i/H$ are relatively insensitive to the applied field up to the
first-order parallel critical field
located approximately half way between the supercooling field (arrow in figure) and
the superheating one (used for normalization in the upper panel). In contrast the gap $\Delta$, while
remaining finite, decreases linearly as expected because
even a small spin-orbit scattering mixes the densities
of states of opposite spins, which in turn are shifted in opposite directions by the Zeeman field.
 The curves are computed by numerically
solving the mean-field equations mentioned in Sec.~\ref{sec:phase};
for concreteness, we use for the spin-orbit scattering parameter
$b=0.052$, and for the Fermi-liquid parameter $G^0=0.155$, as these
values are the one we extract from the measurements in thicker films.
Shown in the inset of the
upper panel of Fig.~\ref{OrderParameter} is the measured parallel
critical field of this film as a function of temperature.  This plot
represents the classic S-P phase diagram in
which a high-temperature line of second-order phase transitions
crosses over into a line of first-order transitions at the tricritical
point $T_{tri}\sim0.3T_c$.\cite{Wu1994}

In the lower panel of Fig.~\ref{OrderParameter} we show the
corresponding behavior of $\delta$, $\Delta$, and $H_i$ at 60~mK for a
7-nm-thick Al film having $T_c\simeq 2.1$~K, $\Delta_o/e \simeq 0.32$~mV,
$\xi\simeq 60$~nm,\cite{footnote1} and $R=16\,\Omega$.  Not only does this thicker film have a
lower critical field due to its finite orbital response, but also the
structure of its phase diagram (lower inset) is dramatically altered
from that of the classic S-P diagram. Indeed, due to the local maximum in
the parallel critical field near $T/T_c=0.45$, the critical-field
behavior is reversibly reentrant.\cite{Wu2006}   The
most obvious difference between the field dependencies shown in
the upper and lower panels of Fig.~\ref{OrderParameter} is that the
gap energy in the thicker film goes to zero before the critical
field is reached.  In fact, the region to the right of the vertical
dot-dashed line represents a gapless superconducting state; clearly,
in this region rapid changes in $\delta$ and $H_i$ take place.

Deep in the superconducting phase, where the superconducting order parameter is well
established, exchange effects are negligible due to the fact that
the quasiparticle density is low and the $e$-$e$ interaction
effects parametrized by  $G^0$ are preempted by the formation of the
condensate.  Interestingly, however, it is clear in both panels
of Fig.~\ref{OrderParameter} that $H_i/H$ never reaches unity in
the limit of small applied field.   We believe that this is a spin-orbit
effect, and that, in the zero-temperature zero-field limit $H_i/H =
1/\left[1+G^0 \alpha(b)\right]$, where $\alpha(b)$ is a positive function of the
spin-orbit parameter $b$.  $\alpha(b)$ goes linearly
to zero as $b\to 0$ and saturates to 1 in the limit
$b\rightarrow \infty$.  Once the field is increased beyond the onset
of the gapless state, the quasiparticle density rapidly grows until it reaches its normal-state
value at $H_{c||}$ as we discuss in more detail below.
Similarly, exchange effects, as reflected in the internal field,
also begin to ``turn on'' upon entering the gapless state.  In
contrast, in the purely S-P limit, there is no gapless state, and the
internal field  jumps discontinuously to its normal-state value at
the first-order critical-field transition (see the upper panel of
Fig.~\ref{OrderParameter}).

\begin{figure}
\includegraphics[width=.48\textwidth]{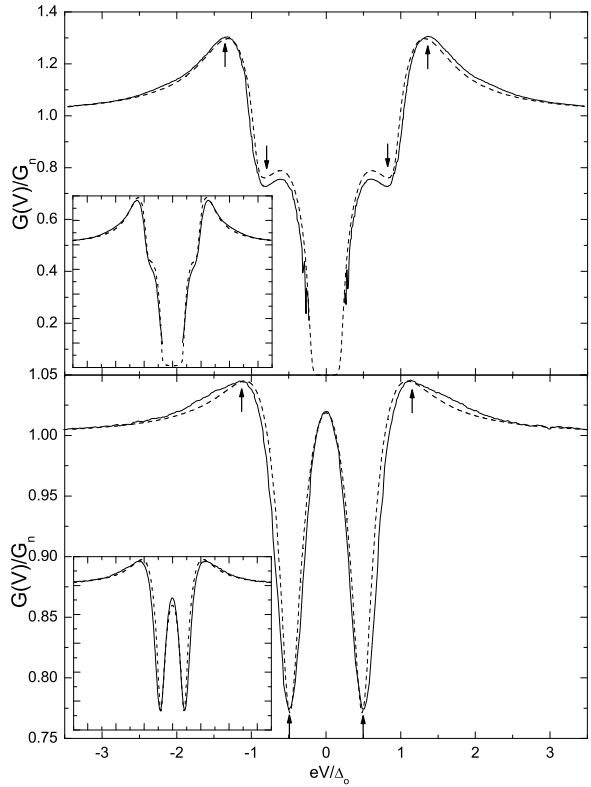}
\caption{\label{TunnelSpectra} Tunneling conductance normalized by
the normal-state conductance $G_n$ as a function of normalized bias voltage.
The solid lines are experimental spectra of a 7-nm-thick Al film taken at 60 mK in a
parallel field of 2.20 (upper panel, gapped) and a 3.02~T (lower panel, gapless).
The dashed curves are the calculated spectra -- see the text for more details.
The arrows point to the primary peak and dip features of the
spectra. The insets show spectra for an 8-nm-thick film taken at 1.40 (upper inset)
and 2.42~T (lower one); the axes cover the same ranges as the respective main panels. The dashed lines
are the theoretical curves hardly distinguishable from the continuous experimental lines.}
\end{figure}

In Fig.~\ref{TunnelSpectra} we compare the normalized tunneling
conductances obtained in the gapped and gapless states of the 7~nm Al
film at 60~mK.  The upper panel was taken at $H=2.2$~T
corresponding to the gapped phase.  The curve clearly displays a
Zeeman splitting of the BCS density-of-states peaks.\cite{Meservey1970}
The data in the lower panel of Fig.~\ref{TunnelSpectra} were
measured at 3.02~T.  Note that the gap is
completely suppressed  and that a large number of states exist at
the Fermi energy ({\it i.e.}, V=0).   The dashed lines in
Fig.~\ref{TunnelSpectra} are the theoretical curves calculated with parameters
$c=0.79$, $G^0=0.155$, and $b=0.052$.\cite{Wu2006}
We also find good agreement between theory and experiment for TDOS measurements
(see insets of Fig.~\ref{TunnelSpectra}) in an
8-nm-thick film with $c\simeq 1.23$; this value is in reasonable agreement
with the scaling $c\propto t^3$ of \eref{Orbital} in the non-local limit
relevant to these samples.\cite{Wu2006}
In the lower panel, an additional broadening,\cite{Dynes} $\Gamma =0.016\Delta_o\simeq
5\times 10^{-6}$~eV,
is used in calculating the theoretical curve; {\it i.e.}, the relation between
density of states $\nu$ and Green's function ${\cal G}$ is taken to be
$\nu(\epsilon)\propto \Im\,{\cal G}(\epsilon+i\Gamma)$.
A broadening of similar magnitude ($\Gamma = 0.01\Delta_o$) is used for the calculated curves
in both insets.
This finite broadening is larger than values reported in the literature,\cite{pek} but at these
fields it could be caused by a small misalignment of the sample. Indeed, the Cooper pair
cyclotron frequency associated with a field misalignment of angle $\theta$ with respect to the sample
plane is $\hbar\Omega_{c\perp}=4eDH\sin\theta$; at $\theta=0.1^\circ$ and
$H= 3$~T, and using the estimate $D\simeq 25$~cm$^2$/s, we obtain
$\hbar\Omega_{c\perp} \simeq 5\times10^{-5} \mathrm{eV}> \Gamma$. We also note that
small variations in the parameters $c$ and $\Gamma$ can compensate each other. On one hand this
seems to support the idea that in the present case $\Gamma$ could be caused by an orbital effect of the
field as $c$ is; on the other hand this interdependence makes it difficult to
extract $\Gamma$ from the data \cite{footnote2} and a more thorough
study of the broadening, including low-field measurements and/or intentional tilting, is necessary
to verify this interpretation.

\begin{figure}
\includegraphics[width=.48\textwidth]{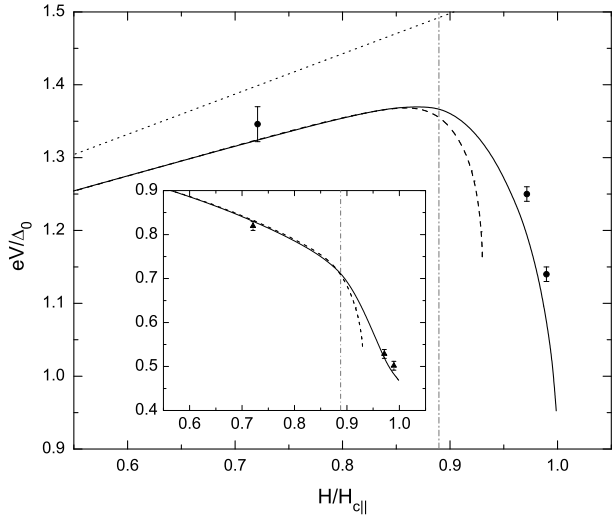}
\caption{\label{PeaksDips} Position of the peaks in Fig.~\ref{TunnelSpectra}
as a function of parallel field. The solid line
represents the theoretical peak positions with Fermi-liquid effects
included. Note that in the gapless state (right of the dot-dashed vertical line), the
peaks move to lower energy as the field increases; this behavior is
in agreement with the experimental measurements (circles with error bars).
The dashed curve shows the theoretical peak positions
in the case of no exchange, {\it i.e.}, $G^0=0$.  The dotted line
represents the Zeeman energy, which is the expected peak
position with no exchange and no orbital pair breaking. Inset: Dip
positions as a function of field.  The solid line is the theoretical
position including exchange, and the dashed line with exchange excluded.
Triangles are data points.}
\end{figure}

Although the two tunneling spectra in Fig.~\ref{TunnelSpectra} are very
different in structure, the evolution
of the spectra as one moves from the gapped to the gapless state is
continuous.  The salient peak and dip features of each spectra are
denoted by the arrows in Fig.~\ref{TunnelSpectra} (we point out that the small broadening
$\Gamma$, introduced above, affects the amplitude but not the position of these features
in the theoretical curves).
The behavior of these features as one crosses over into the gapless
state is shown in Fig.~\ref{PeaksDips}.  The main panel depicts the
low-temperature field dependence of the outermost coherence peaks
along with data points corresponding to tunneling measurements. Interestingly, the
position of the peaks is not monotonic in field, and the
peaks move to lower energy in the gapless state.  The
dotted line is simply the Zeeman splitting, which
represents the expected peak position if there were no orbital
depairing and no exchange effects.
We have verified numerically that, in contrast to the results for the marginally thin films,
the peaks follow the bare Zeeman splitting with good approximation
for the thinner film with $c=0.02$, in agreement with low-temperature
measurements in thinner films.\cite{Meservey1970}
The dashed curve is the
expected peak position with orbital depairing but no exchange. This
curve misses the data points badly while agreeing with the solid
curve at low field indicating that $G^0$ plays a
significant role in determining the tunneling density of states in the gapless
state. Moreover, with $G^0=0$ the low-temperature phase transition
is predicted to be first-order, in contrast to the experimental finding
of a second-order transition.
The inset of Fig.~\ref{PeaksDips} shows the corresponding
field dependence of the spectra dips. As is the case with the peaks, the evolution of the dips' position
contrasts with that of the thin-film case. In the latter, the dips follow the
peaks to higher energies. In Fig.~\ref{PeaksDips},
the dips move to {\it lower} energy and the peaks broaden with increasing field.
This broadening could potentially limit the degree of
spin polarization that could be achieved in hybrid structures.\cite{Giaz}

\begin{figure}
\includegraphics[width=.48\textwidth]{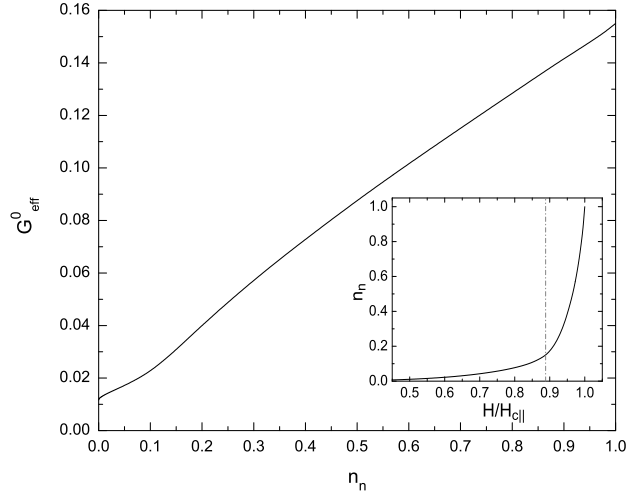}
\caption{\label{Exchange} Effective FL parameter $G^0_\mathrm{eff}$ [\eref{Geff}] as a function of
quasiparticle density $n_n$ in the superconducting phase of a marginally thin Al film.
The inset is the calculated quasiparticle density as a function of
the normalized parallel magnetic field for a 7-nm-thick Al film at 60 mK.  The
vertical dot-dashed line divides the gapped and gapless states. Note that in the latter region,
the quasiparticle density quickly rises with the applied field.}
\end{figure}

Building on the above analysis we can now extract an ``effective'' FL
exchange parameter $G^{0}_\mathrm{eff}$ as a function of the
quasiparticle density $n_n$.  In Fig.~\ref{Exchange}
we plot such a function, where we assume a two-fluid
model in order to estimate $n_n$. In this phenomenological approach, the superconducting carrier
density $n_{s}$ is related to the penetration depth $\lambda$ by
\begin{equation}
n_s \propto \lambda^{-2}
\end{equation}
and the normal electron density is $n_{n}=1-n_{s}$.
In disordered thin superconducting films the parallel field
penetration depth is proportional (up to a weakly temperature dependent coefficient) to the
penetration depth $\lambda_{L}$ of a clean superconductor in the local limit [see, e.g., Eq. (3.136) of
\ocite{Tinkham1996}]. The latter depends on temperature approximately as
\begin{equation}\label{lambda}
\lambda_L(\gamma) \propto \frac{1}{\sqrt{1-\gamma^2}}\, ,
\end{equation}
where $\gamma=T/T_c$. Near the critical temperature, the temperature dependence of
the order parameter is $\delta(\gamma)/\Delta_0 \propto \sqrt{1-\gamma}$.
Inverting this relationship to express $\gamma$ in terms of $\delta$, substituting the result into
Eq.~(\ref{lambda}), and using the definition of $n_n$, we arrive at:
\begin{equation}
n_n \propto \left(1-\frac{\delta^2}{\Delta_0^2}\right)^2.
\end{equation}
From this relation we evaluate the field dependence of the normal
electron density through the calculated field dependence of the
order parameter (at low temperature the proportionality constant can
be taken to be unity up to exponentially small corrections); the
result is presented in the inset of Fig.~\ref{Exchange}.  We define the
effective Fermi-liquid constant by
\be\label{Geff}
G^0_\mathrm{eff} = H/H_i-1 \ .
\ee
We can obtain this quantity from the
calculated ratio $H_i/H$ shown in Fig.~\ref{OrderParameter}.  Note
that $G^{0}_\mathrm{eff}$ is approximately linear in the quasiparticle
density, thus showing the gradual rise of exchange effects going from the
superconducting to the normal phase. This microscopic calculation
fully supports the qualitative explanation given in \ocite{Butko1999} for different g-factors in the
superconducting and normal states, which correspond to $G^0_\mathrm{eff}=0$ and
$G^0_\mathrm{eff}\simeq0.17$, respectively.

In summary, we have presented measurements of the tunneling density of states of marginally
thin Al films in parallel magnetic field. The marginality condition
$c\sim 1$, with the dimensionless parameter $c$ defined in \eref{Orbital},
gives access to a gapless superconducting state.
The inclusion of Fermi-liquid effects, parametrized by the quantity $G^0\sim 0.16$, is
necessary to explain the main features (peaks and dips) of the tunneling spectra, see
Figs.~\ref{TunnelSpectra} and \ref{PeaksDips}. Although these interaction effects are suppressed deep
in the superconducting phase, they become more and more important as one moves through the gapless
state to the parallel critical-field transition.  In the gapless region the effective
Fermi-liquid parameter $G^0_\mathrm{eff}$ grows quasi-linearly with the quasiparticle density.

\acknowledgments

We gratefully  acknowledge enlightening discussions with Ilya Vekhter and Dana
Browne. P.W.A. acknowledges the support of the DOE under Grant No.\ DE-FG02-07ER46420.


\begin{thebibliography}{30}
\bibitem{footnote2}For example, no broadening is used in the upper panel curve -- including it
would slightly suppress the peaks' magnitude. Due to their interdependence, however, different values
for $c$ and $\Gamma\neq 0$ could also satisfactorily fit the data.
\end{thebibliography}
\end{document}